\documentclass[%
 reprint,
superscriptaddress,
 amsmath,amssymb,
 aps,
 prstab,
 letterpaper,
 ]{revtex4-1}

\usepackage[english]{babel}
\usepackage[utf8x]{inputenc}
\usepackage[T1]{fontenc}

\usepackage[a4paper,top=3cm,bottom=2cm,left=2cm,right=2cm,marginparwidth=1.75cm]{geometry}

\usepackage{amsmath}
\usepackage{units}
\usepackage{graphicx}
\usepackage[colorinlistoftodos]{todonotes}
\usepackage[colorlinks=true, allcolors=blue]{hyperref}
\usepackage{setspace}
\usepackage{lineno}
\usepackage{graphicx}
\usepackage{dcolumn}
\usepackage{bm}

\begin{document}
\title{Experimental study of extended timescale dynamics of a plasma wakefield driven by a self-modulated proton bunch}

\author{J.~Chappell}
\affiliation{UCL, London, UK}
\author{E.~Adli}
\affiliation{University of Oslo, Oslo, Norway}
\author{R.~Agnello}
\affiliation{Ecole Polytechnique Federale de Lausanne (EPFL), Swiss Plasma Center (SPC), Lausanne, Switzerland}
\author{M.~Aladi}
\affiliation{Wigner Research Center for Physics, Budapest, Hungary}
\author{Y.~Andrebe}
\affiliation{Ecole Polytechnique Federale de Lausanne (EPFL), Swiss Plasma Center (SPC), Lausanne, Switzerland}
\author{O.~Apsimon}
\affiliation{Cockcroft Institute, Daresbury, UK}
\affiliation{Lancaster University, Lancaster, UK}
\author{R.~Apsimon}
\affiliation{Cockcroft Institute, Daresbury, UK}
\affiliation{Lancaster University, Lancaster, UK}
\author{A.-M.~Bachmann}
\affiliation{CERN, Geneva, Switzerland}
\affiliation{Max Planck Institute for Physics, Munich, Germany}
\affiliation{Technical University Munich, Munich, Germany}
\author{M.A.~Baistrukov}
\affiliation{Budker Institute of Nuclear Physics SB RAS, Novosibirsk, Russia}
\affiliation{Novosibirsk State University, Novosibirsk, Russia}
\author{F.~Batsch}
\affiliation{CERN, Geneva, Switzerland}
\affiliation{Max Planck Institute for Physics, Munich, Germany}
\affiliation{Technical University Munich, Munich, Germany}
\author{M.~Bergamaschi}
\affiliation{CERN, Geneva, Switzerland}
\author{P.~Blanchard}
\affiliation{Ecole Polytechnique Federale de Lausanne (EPFL), Swiss Plasma Center (SPC), Lausanne, Switzerland}
\author{P.N.~Burrows}
\affiliation{John Adams Institute, Oxford University, Oxford, UK}
\author{B.~Buttensch{\"o}n}
\affiliation{Max Planck Institute for Plasma Physics, Greifswald, Germany}
\author{A.~Caldwell}
\affiliation{Max Planck Institute for Physics, Munich, Germany}
\author{E.~Chevallay}
\affiliation{CERN, Geneva, Switzerland}
\author{M.~Chung}
\affiliation{UNIST, Ulsan, Republic of Korea}
\author{D.A.~Cooke}
\affiliation{UCL, London, UK}
\author{H.~Damerau}
\affiliation{CERN, Geneva, Switzerland}
\author{C.~Davut}
\affiliation{Cockcroft Institute, Daresbury, UK}
\affiliation{University of Manchester, Manchester, UK}
\author{G.~Demeter}
\affiliation{Wigner Research Center for Physics, Budapest, Hungary}
\author{L.H.~Deubner}
\affiliation{Philipps-Universit{\"a}t Marburg, Marburg, Germany}
\author{A.~Dexter}
\affiliation{Cockcroft Institute, Daresbury, UK}
\affiliation{Lancaster University, Lancaster, UK}
\author{G.P.~Djotyan}
\affiliation{Wigner Research Center for Physics, Budapest, Hungary}
\author{S.~Doebert}
\affiliation{CERN, Geneva, Switzerland}
\author{J.~Farmer}
\affiliation{CERN, Geneva, Switzerland}
\affiliation{Max Planck Institute for Physics, Munich, Germany}
\author{A.~Fasoli}
\affiliation{Ecole Polytechnique Federale de Lausanne (EPFL), Swiss Plasma Center (SPC), Lausanne, Switzerland}
\author{V.N.~Fedosseev}
\affiliation{CERN, Geneva, Switzerland}
\author{R.~Fiorito}
\affiliation{Cockcroft Institute, Daresbury, UK}
\affiliation{University of Liverpool, Liverpool, UK}
\author{R.A.~Fonseca}
\affiliation{ISCTE - Instituto Universit\'{e}ario de Lisboa, Portugal}
\affiliation{GoLP/Instituto de Plasmas e Fus\~{a}o Nuclear, Instituto Superior T\'{e}cnico, Universidade de Lisboa, Lisbon, Portugal}
\author{F.~Friebel}
\affiliation{CERN, Geneva, Switzerland}
\author{I.~Furno}
\affiliation{Ecole Polytechnique Federale de Lausanne (EPFL), Swiss Plasma Center (SPC), Lausanne, Switzerland}
\author{L.~Garolfi}
\affiliation{TRIUMF, Vancouver, Canada}
\author{S.~Gessner}
\affiliation{CERN, Geneva, Switzerland} 
\affiliation{SLAC National Accelerator Laboratory, Menlo Park, California, USA}
\author{B.~Goddard}
\affiliation{CERN, Geneva, Switzerland} 
\author{I.~Gorgisyan}
\affiliation{CERN, Geneva, Switzerland}
\author{A.A.~Gorn}
\affiliation{Budker Institute of Nuclear Physics SB RAS, Novosibirsk, Russia} 
\affiliation{Novosibirsk State University, Novosibirsk, Russia}
\author{E.~Granados}
\affiliation{CERN, Geneva, Switzerland}
\author{M.~Granetzny}
\affiliation{University of Wisconsin, Madison, Wisconsin, USA}
\author{O.~Grulke}
\affiliation{Max Planck Institute for Plasma Physics, Greifswald, Germany}
\affiliation{Technical University of Denmark, Lyngby, Denmark}
\author{E.~Gschwendtner}
\affiliation{CERN, Geneva, Switzerland} 
\author{V.~Hafych}
\affiliation{Max Planck Institute for Physics, Munich, Germany}
\author{A.~Hartin}
\affiliation{UCL, London, UK}
\author{A.~Helm}
\affiliation{GoLP/Instituto de Plasmas e Fus\~{a}o Nuclear, Instituto Superior T\'{e}cnico, Universidade de Lisboa, Lisbon, Portugal}
\author{J.R.~Henderson}
\affiliation{Cockcroft Institute, Daresbury, UK}
\affiliation{Accelerator Science and Technology Centre, ASTeC, STFC Daresbury Laboratory, Warrington, UK}
\author{A.~Howling}
\affiliation{Ecole Polytechnique Federale de Lausanne (EPFL), Swiss Plasma Center (SPC), Lausanne, Switzerland}
\author{M.~H{\"u}ther}
\affiliation{Max Planck Institute for Physics, Munich, Germany}
\author{R.~Jacquier}
\affiliation{Ecole Polytechnique Federale de Lausanne (EPFL), Swiss Plasma Center (SPC), Lausanne, Switzerland}
\author{S.~Jolly}
\affiliation{UCL, London, UK}
\author{I.Yu.~Kargapolov}
\affiliation{Budker Institute of Nuclear Physics SB RAS, Novosibirsk, Russia} 
\affiliation{Novosibirsk State University, Novosibirsk, Russia}
\author{M.{\'A}.~Kedves}
\affiliation{Wigner Research Center for Physics, Budapest, Hungary}
\author{F.~Keeble}
\affiliation{UCL, London, UK}
\author{M.D.~Kelisani}
\affiliation{CERN, Geneva, Switzerland}
\author{S.-Y.~Kim}
\affiliation{UNIST, Ulsan, Republic of Korea}
\author{F.~Kraus}
\affiliation{Philipps-Universit{\"a}t Marburg, Marburg, Germany}
\author{M.~Krupa}
\affiliation{CERN, Geneva, Switzerland}
\author{T.~Lefevre}
\affiliation{CERN, Geneva, Switzerland}
\author{Y.~Li}
\affiliation{Cockcroft Institute, Daresbury, UK}
\affiliation{University of Manchester, Manchester, UK}
\author{L.~Liang}
\affiliation{Cockcroft Institute, Daresbury, UK}
\affiliation{University of Manchester, Manchester, UK}
\author{S.~Liu}
\affiliation{TRIUMF, Vancouver, Canada}
\author{N.~Lopes}
\affiliation{GoLP/Instituto de Plasmas e Fus\~{a}o Nuclear, Instituto Superior T\'{e}cnico, Universidade de Lisboa, Lisbon, Portugal}
\author{K.V.~Lotov}
\affiliation{Budker Institute of Nuclear Physics SB RAS, Novosibirsk, Russia}
\affiliation{Novosibirsk State University, Novosibirsk, Russia}
\author{M.~Martyanov}
\affiliation{Max Planck Institute for Physics, Munich, Germany}
\author{S.~Mazzoni}
\affiliation{CERN, Geneva, Switzerland}
\author{D.~Medina~Godoy}
\affiliation{CERN, Geneva, Switzerland}
\author{V.A.~Minakov}
\affiliation{Budker Institute of Nuclear Physics SB RAS, Novosibirsk, Russia}
\affiliation{Novosibirsk State University, Novosibirsk, Russia}
\author{J.T.~Moody}
\affiliation{Max Planck Institute for Physics, Munich, Germany}
\author{P.I.~Morales~Guzm\'{a}n}
\affiliation{Max Planck Institute for Physics, Munich, Germany}
\author{M.~Moreira}
\affiliation{CERN, Geneva, Switzerland}
\affiliation{GoLP/Instituto de Plasmas e Fus\~{a}o Nuclear, Instituto Superior T\'{e}cnico, Universidade de Lisboa, Lisbon, Portugal}
\author{H.~Panuganti}
\affiliation{CERN, Geneva, Switzerland} 
\author{A.~Pardons}
\affiliation{CERN, Geneva, Switzerland}
\author{F.~Pe\~na~Asmus}
\affiliation{Max Planck Institute for Physics, Munich, Germany}
\affiliation{Technical University Munich, Munich, Germany}
\author{A.~Perera}
\affiliation{Cockcroft Institute, Daresbury, UK}
\affiliation{University of Liverpool, Liverpool, UK}
\author{A.~Petrenko}
\affiliation{Budker Institute of Nuclear Physics SB RAS, Novosibirsk, Russia}
\author{J.~Pucek}
\affiliation{Max Planck Institute for Physics, Munich, Germany}
\author{A.~Pukhov}
\affiliation{Heinrich-Heine-Universit{\"a}t D{\"u}sseldorf, D{\"u}sseldorf, Germany}
\author{B.~R\'{a}czkevi}
\affiliation{Wigner Research Center for Physics, Budapest, Hungary}
\author{R.L.~Ramjiawan}
\affiliation{CERN, Geneva, Switzerland}
\affiliation{John Adams Institute, Oxford University, Oxford, UK}
\author{S.~Rey}
\affiliation{CERN, Geneva, Switzerland}
\author{H.~Ruhl}
\affiliation{Ludwig-Maximilians-Universit{\"a}t, Munich, Germany}
\author{H.~Saberi}
\affiliation{CERN, Geneva, Switzerland}
\author{O.~Schmitz}
\affiliation{University of Wisconsin, Madison, Wisconsin, USA}
\author{E.~Senes}
\affiliation{CERN, Geneva, Switzerland}
\affiliation{John Adams Institute, Oxford University, Oxford, UK}
\author{P.~Sherwood}
\affiliation{UCL, London, UK}
\author{L.O.~Silva}
\affiliation{GoLP/Instituto de Plasmas e Fus\~{a}o Nuclear, Instituto Superior T\'{e}cnico, Universidade de Lisboa, Lisbon, Portugal}
\author{R.I.~Spitsyn}
\affiliation{Budker Institute of Nuclear Physics SB RAS, Novosibirsk, Russia}
\affiliation{Novosibirsk State University, Novosibirsk, Russia}
\author{P.V.~Tuev}
\affiliation{Budker Institute of Nuclear Physics SB RAS, Novosibirsk, Russia}
\affiliation{Novosibirsk State University, Novosibirsk, Russia}
\author{F.~Velotti}
\affiliation{CERN, Geneva, Switzerland}
\author{L.~Verra}
\affiliation{CERN, Geneva, Switzerland}
\affiliation{Max Planck Institute for Physics, Munich, Germany}
\author{V.A.~Verzilov}
\affiliation{TRIUMF, Vancouver, Canada} 
\author{J.~Vieira}
\affiliation{GoLP/Instituto de Plasmas e Fus\~{a}o Nuclear, Instituto Superior T\'{e}cnico, Universidade de Lisboa, Lisbon, Portugal}
\author{C.P.~Welsch}
\affiliation{Cockcroft Institute, Daresbury, UK}
\affiliation{University of Liverpool, Liverpool, UK}
\author{B.~Williamson}
\affiliation{Cockcroft Institute, Daresbury, UK}
\affiliation{University of Manchester, Manchester, UK}
\author{M.~Wing}
\affiliation{UCL, London, UK}
\author{J.~Wolfenden}
\affiliation{Cockcroft Institute, Daresbury, UK}
\affiliation{University of Liverpool, Liverpool, UK}
\author{B.~Woolley}
\affiliation{CERN, Geneva, Switzerland}
\author{G.~Xia}
\affiliation{Cockcroft Institute, Daresbury, UK}
\affiliation{University of Manchester, Manchester, UK}
\author{M.~Zepp}
\affiliation{University of Wisconsin, Madison, Wisconsin, USA}
\author{G.~Zevi~Della~Porta}
\affiliation{CERN, Geneva, Switzerland}
\collaboration{The AWAKE Collaboration}
\noaffiliation

\sloppy 

\begin{abstract}
    Plasma wakefield dynamics over timescales up to 800\,ps, approximately 100 plasma periods, are studied experimentally at the Advanced Wakefield Experiment (AWAKE). The development of the longitudinal wakefield amplitude driven by a self-modulated proton bunch is measured using the external injection of witness electrons that sample the fields. In simulation, resonant excitation of the wakefield causes plasma electron trajectory crossing, resulting in the development of a potential outside the plasma boundary as electrons are transversely ejected. Trends consistent with the presence of this potential are experimentally measured and their dependence on wakefield amplitude are studied via seed laser timing scans and electron injection delay scans.
\end{abstract}

\maketitle

The Advanced Wakefield Experiment (AWAKE) is a proof-of-principle plasma wakefield acceleration (PWFA) experiment that has demonstrated the acceleration of an externally injected witness bunch in the wakefield driven by a self-modulated proton bunch~\cite{caldwellawake, nature}. The use of a high-energy proton bunch as a PWFA drive beam offers the potential for acceleration of electron beams to TeV-scale energies in a single plasma stage~\cite{tevwitness} thanks to the large stored energy of the drive bunch (e.g.\,for an LHC proton bunch, $E \sim 130$\,kJ), more than three orders of magnitude greater than the energy stored in bunches typically used in electron-driven PWFA experiments~\cite{flashforward,facet2}. 

In order to optimally excite the wakefield, drive bunches used in PWFA experiments must be short with respect to the plasma wavelength. The bunch RMS length, $\sigma_z$, should ideally satisfy the relation $\sigma_z < \lambda_p /2 = \pi c / \omega_p$ where $\omega_p = \sqrt{n_e e^2 / m_e \varepsilon_0}$ is the electron angular plasma frequency, $n_e$ is the electron plasma density, $e$ and $m_e$ are the charge and mass of the electron, $\varepsilon_0$ is the vacuum permittivity, and $c$ is the speed of light. The Super Proton Synchrotron (SPS) proton bunch used by AWAKE typically varies in RMS length between 6 -- 8\,cm and hence is much longer than required for plasma densities of interest ($n_e \sim 10^{15}$\,cm$^{-3}$ gives $\lambda_p \sim 1$\,mm and a cold wave-breaking field in excess of 3\,GVm$^{-1}$). However, it is possible to take advantage of a natural response when a plasma is perturbed by a long ($\sigma_z \gg \lambda_p$) charged particle bunch: self-modulation (SM). Low amplitude transverse wakefields driven by the long proton bunch focus and defocus alternating regions of the bunch. This causes modulation into micro-bunches that can each effectively drive a wakefield within the plasma~\cite{selfmodulation}. Due to this mechanism being an intrinsic plasma response, the micro-bunches are naturally separated by the plasma wavelength allowing resonant excitation of the wakefield~\cite{ssm1,ssm2}. The cumulative effect of the wakefield driven by multiple micro-bunches can result in longitudinal wakefields with amplitudes reaching GVm$^{-1}$~\cite{tevwitness}, more than an order of magnitude larger than fields that are typically achieved in traditional radio-frequency (RF) accelerating cavities. 

It is possible to seed the development of SM by driving sufficiently strong initial transverse wakefields. This ensures SM is the dominant evolution mechanism rather than competing non-axisymmetric modes such as hosing~\cite{smvshosing}. When SM is seeded it is expected that the phase of the wakefield is reproducible, essential for consistent acceleration of externally injected short witness bunches. At AWAKE this is achieved by co-propagating the ionising laser pulse close to the centre of the proton bunch. This creates a sharp plasma ionisation front that interacts with a high density region of the bunch, inducing large amplitude ($>$\,MVm$^{-1}$) seed wakefields~\cite{awakereadiness}. 

In this study, the extended timescale evolution of a plasma wakefield driven by a self-modulated proton bunch is experimentally probed. Measurements of the integrated longitudinal wakefield by acceleration of externally injected witness electrons demonstrate resonant excitation of the wakefield along the self-modulated proton bunch until phase mixing of plasma electrons~\cite{dawsonmixing} causes saturation of the wakefield amplitude. Simulations correspondingly predict the development of a potential well that is attractive for electrons outside the plasma boundary on extended timescales due to the subsequent ejection of plasma electrons. These predictions are consistent with experimental observations of decreasing witness energy and increasing captured charge with increasing witness injection delay. The amplitude, and hence influence, of this potential is studied by changing the plasma wakefield amplitude.

Understanding the evolution of the wakefield amplitude along both the proton bunch and the plasma is essential for optimisation of the acceleration process for future applications~\cite{awakeapplications2}. Experimentally this is measured by varying the temporal delay between the seeding laser pulse and the injection of a witness electron bunch that samples the wakefield. This is henceforth referred to as the \textit{laser--electron delay} and permits measurement of the integrated longitudinal field local to the position of the captured electrons along the self-modulated proton bunch. In addition, the maximum wakefield amplitude driven by the self-modulated bunch can be varied by adjusting the relative time of arrival of the seeding laser pulse and the proton bunch, henceforth referred to as the \textit{seeding position}. Moving the seeding position within the proton bunch changes the fraction of the bunch that interacts with the plasma and hence the number of protons that drive the wakefield experienced by the witness electrons.

\begin{figure*}[t]
	\centering
	\includegraphics[width=\textwidth]{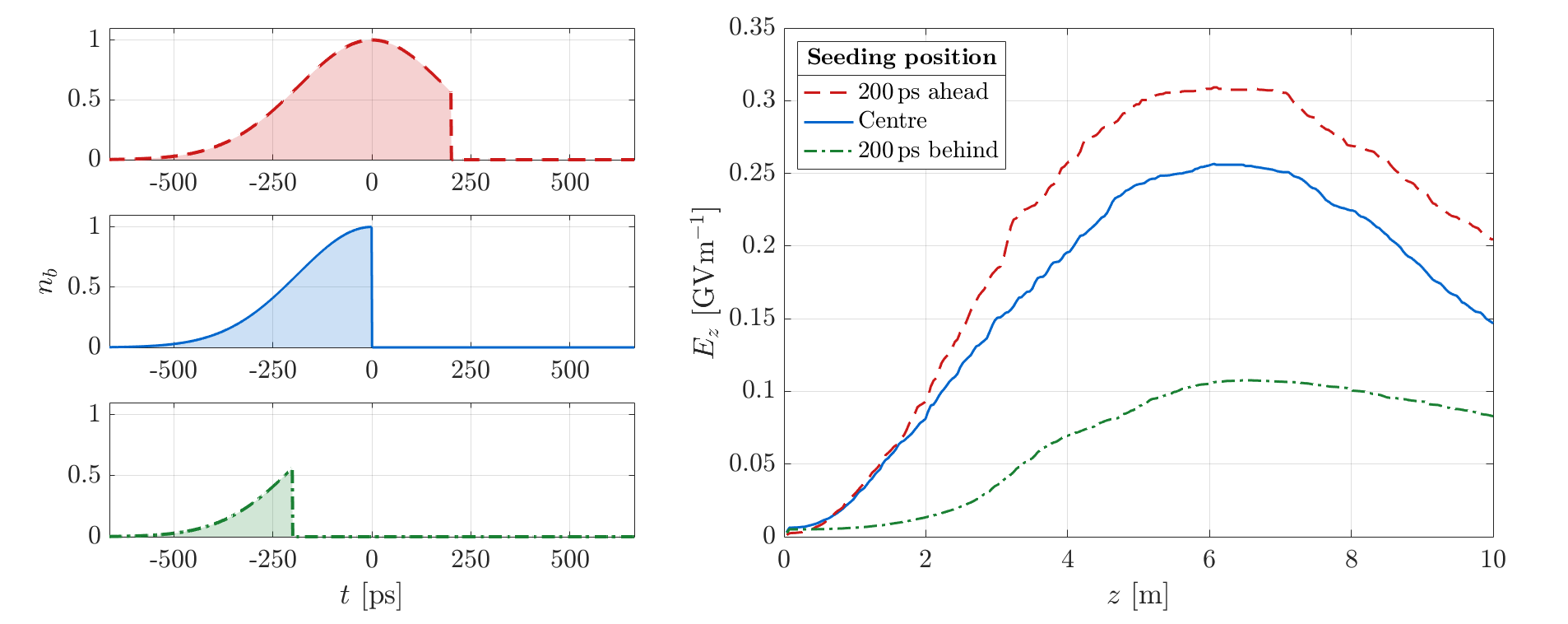}
    \caption{Left: Illustration of the normalised beam density, $n_b$, corresponding to the seeding positions used in this study. Shaded regions indicate the region of the beam that propagates through, and hence interacts with, the plasma. Left, upper: Seeding 200\,ps ahead of the centre. Left, middle: Seeding in the centre of the proton bunch. Left, lower: Seeding 200\,ps behind the centre. Right: Simulated evolution of the maximum longitudinal field over the entire plasma length for the three different seeding positions for a plasma density of $n_e = 2 \times 10^{14}\,$cm$^{-3}$. The proton bunch parameters are the same as those used for studying extended timescale wakefield evolution described in Sec.~\ref{sec:longtimescale}.}
    \label{fig:beamdenscomp}
\end{figure*}

This study discusses results obtained during two experimental periods of AWAKE Run 1 in September (Sec.~\ref{sec:resonantexcitation}) and November (Sec.~\ref{sec:longtimescale}) 2018. The experiments involved varying the laser--electron delay, $\Delta \tau$, from 50\,ps to 800\,ps for three different seeding positions: in the centre of the proton bunch ($t = 0\,$ps) where the initial bunch density interacting with the plasma, and hence amplitude of the seed wakefields, are highest; and 200\,ps ahead of, or behind, the centre of the proton bunch as shown in the left panel of Fig.~\ref{fig:beamdenscomp}. In these latter cases the seed wakefields are similar but smaller in magnitude than those when seeding in the centre, while the proton bunch density is initially increasing, or decreasing, along the bunch, respectively resulting in a change in the evolution of the longitudinal wakefield amplitude. Figure~\ref{fig:beamdenscomp} shows that the maximal longitudinal wakefield amplitude increases with the number of protons driving wakefields.

\section{Experimental overview}

The proton bunch was extracted from the SPS and transported to the experimental area where it was focused to a transverse size of $\sigma_r \approx 200\,\mathrm{\mu}$m at the entrance of the vapour source and had a transverse emittance of approximately 3.5\,mm$\cdot$mrad. A 300\,pC witness electron bunch was generated using a frequency-tripled derivative of the main ionising laser pulse to ensure timing stability at the picosecond-level between the electron bunch and seeding laser pulse. The electron bunch was accelerated to 19\,MeV using traditional RF cavities and transported along the beam line to the entrance of the vapour source where it is injected into the wakefields driven by the self-modulated proton bunch~\cite{electronsource}. The electron bunch trajectory was matched to that of the proton bunch with a small vertical offset and injected into the plasma at an angle of approximately 0.5\,mrad. It was focused to an RMS transverse size of approximately 500\,$\mu$m at the entrance of the vapour source with a bunch length similar to the plasma wavelength, $\sigma_z \approx 8\,\mathrm{ps}$. Portions of the witness bunch are hence captured within multiple accelerating wakefield buckets and therefore it was not possible to extract detailed longitudinal field structure dependencies below this temporal limit. The relative timing of the ionising seeding laser pulse and witness electron bunch was adjusted using a delay stage in the transport line of the laser pulse to the photocathode. Two quadrupoles were placed 4.48 and 4.98\,m downstream of the exit of the plasma to capture and focus the accelerated witness electrons before they were horizontally dispersed by a 1\,m-long C-shaped electromagnetic dipole and imaged on a scintillator screen~\cite{spectrometer}. Light emitted by the scintillator screen was imaged onto an intensified CCD camera. The relationship between the position of an electron in the plane of the scintillator screen and its energy was calculated using the Beam Delivery Simulation (BDSIM) code~\cite{bdsim,geant4} with measured dipole field maps as input. The energy uncertainty was approximately 2\,\%, calculated via considerations of the accuracy of the field maps, measurements of the positions of the spectrometer beamline components, and the resolution of the spectrometer imaging system~\cite{fearghus}. For the measurements presented hereafter, the dipole current was kept constant at 100\,A, allowing electron energies ranging from 84\,MeV to 2\,GeV to be measured. 

Calibration of the charge response of the scintillator screen was performed and validated via two independent methods. Firstly, electron beams of variable charge were used to study the scintillator response at the CERN Linear Electron Accelerator for Research (CLEAR) test beam facility~\cite{spectrometer}. Secondly, beams of mono-energetic electrons were produced by the stripping of high-energy lead ions accelerated in the SPS and transported to the AWAKE experimental area to be imaged by the spectrometer system in-situ~\cite{psi}. These two complementary measurements permitted calculation of the witness bunch charge from measurements of the integrated light output from the scintillator with an associated uncertainty of 8\,\%.

The vapour is contained in a 10\,m-long cylindrical cell of diameter 40\,mm with Rb reservoirs at either end~\cite{ozvapoursource}. The cell was heated to provide tunable vapour density. For these studies, the vapour density at both ends was kept equal and constant at a value of $2 \times 10^{14}$\,cm$^{-3}$. The density of the vapour was monitored by an interferometric measurement at each end of the cell using white light interferometry around the 780 and 795\,nm lines of the Rb atom, with an uncertainty of 0.5\,\%~\cite{rbdensity}. While larger witness energies could be observed at higher plasma densities~\cite{nature}, achieving consistent witness capture was more challenging and hence lower operating densities were preferred for systematic studies.

Ionisation of the Rb vapour was achieved using a terawatt-class Ti:sapphire laser. The laser pulse duration was approximately 120\,fs with a pulse energy that can be varied from 40 to 450\,mJ. The Rayleigh length of the focused laser pulse was 15\,m, with a spot size of approximately 2\,mm throughout the entire vapour source. It was assumed that the laser singly-ionised the Rb vapour in accordance with previous measurements that demonstrated self-modulation of the proton bunch at a frequency consistent with that of the measured vapour density~\cite{ssm1}. The phase of the laser oscillator was locked to the radio frequency of the cavities within the SPS, with synchronisation between the laser pulse and time of arrival of the proton beam measured to be at the picosecond level, thus permitting controlled variations of the seed position.  

\subsection*{Analysis of spectrometer images}

For an event to be included in this analysis, a minimum witness charge of 50\,fC needed to be detected on the spectrometer screen in order to eliminate events where sufficient witness capture was not achieved. For each event, background subtraction and geometric corrections~\cite{fearghus} were applied to the image of the scintillator screen and the pixel count was integrated over the vertical (non-dispersive) plane. The region with signal that exceeded the expected background by 3\,$\sigma_{bkg}$ was identified. The mean energy of the captured witness electrons was calculated according to $\mu_E = \left(\Sigma_i\, E_i \cdot dQ_i \right) / \left(\Sigma_i\, dQ_i \right)$ where $i$ corresponds to the index of the column of the image in the signal region, $E_i$ is the energy associated with column $i$ in the plane of the scintillator screen and $dQ_i$ is the integrated charge measured in column $i$. The integrated charge of a column was calculated by summing the total CCD pixel counts of the background-subtracted image and applying the calibrated scintillator charge response value, $\left( 4.22 \pm 0.33 \right) \times 10^5$\,pC$^{-1}$.

\section{Simulation overview}

The experimental measurements are compared to 2D cylindrical, quasi-static simulations performed using LCODE~\cite{lcode}. These simulations solve for the plasma response in the co-moving frame, defined by $\xi = z - ct$, and use the experimental proton and electron bunch parameters as input. The simulation domain spans $0 \leq \xi \leq 800\,k_p^{-1}$, $0 \leq r \leq 25\,k_p^{-1}$ with a resolution of $\Delta \xi = \Delta r = 0.02\,k_p^{-1}$ where $k_p = \omega_p / c$. A fixed timestep of $\Delta t = 100\,\omega_p^{-1}$ is used to update both the plasma state and proton beam; for witness electrons, an energy-dependent reduced timestep is used to fully resolve their betatron oscillations. Approximately 30 radius-weighted macro-particles per cell per species are used to model the response of plasma electrons and ions with $2.7\times10^6$ equal-weighted beam macro-particles used to model the proton bunch. Use of a wide simulation domain that extends far beyond the boundary of the plasma ($r_p = 5.3\,k_p^{-1} = 2$\,mm) is necessitated by the requirement to track plasma electrons that are expelled from the plasma after trajectory crossing occurs as is discussed in more detail later and in Ref.~\cite{gorndefocusing}. The proton bunch is initialised with Gaussian longitudinal and radial distributions. Modelling of the varying seeding positions used in this study is achieved by using a step function in the proton bunch density at the relevant seeding position within the bunch. This negates the need to model the interaction between the ionising laser pulse and the Rb vapour and hence saves computational resources. 

The process of external electron injection is inherently a complicated 3D problem~\cite{3dpicinjection} and therefore perfect agreement between simulations and experiment is not expected. In 2D cylindrical simulations, the witness electrons are represented by a ring of charge being injected into the wakefield toward the axis as opposed to an electron bunch with pointing jitters as in the experiment. As such, the total captured charge in 2D is expected to exceed that observed experimentally. $\mathcal{O}$(pC) captured witness charges were typically measured during the experiment and hence no significant modification of the wakefield amplitude experienced by the witness electrons via beam loading is expected. For these two reasons, test witness electrons that can experience the wakefield but do not alter its amplitude or drive their own wake are used in simulation in an attempt to recreate measured experimental trends. 

It is additionally expected that 2D axisymmetric simulations underestimate turbulent effects due to the imposed symmetry and hence likely provide an overestimate of the wakefield amplitude at large laser--electron delays following phase mixing. However, at present, three-dimensional quasi-static simulations of the extended plasma region required for comparison to experimental observations are prohibitively computationally expensive. In spite of this, axisymmetric simulations still provide useful comparison and insight into the underlying physical mechanisms.


\begin{figure}[b]
	\centering
	\includegraphics[width=0.49\textwidth]{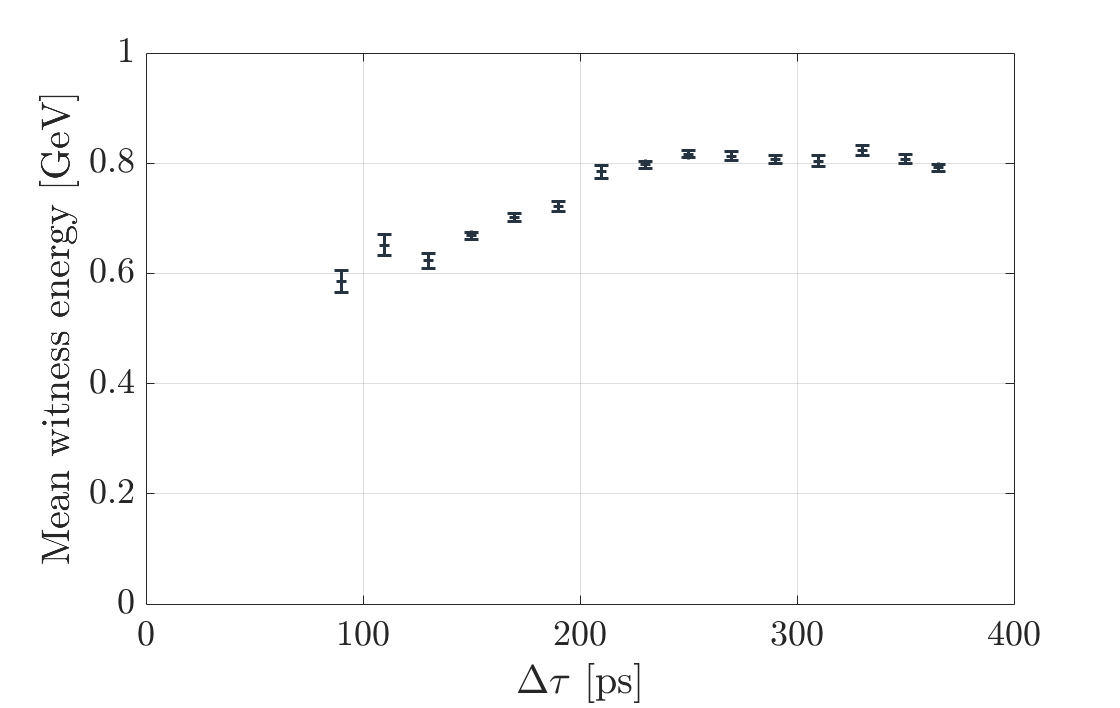}
    \caption{Experimental results showing mean witness energy for $\Delta \tau$ up to 365\,ps when seeding 100\,ps ahead of the centre of the proton bunch. The growth in mean witness energy with increasing $\Delta \tau$ is consistent with resonant excitation of the longitudinal wakefield. Error bars represent the standard error on the mean: $\sigma({\mu_E})/\sqrt{N-1}$, where $\sigma(\mu_E)$ is the standard deviation of the measurements of the mean energy and $N$ represents the number of events per step.}
    \label{fig:shortdelays}
\end{figure}

\begin{figure*}[t]
	\centering
	\includegraphics[width=\textwidth]{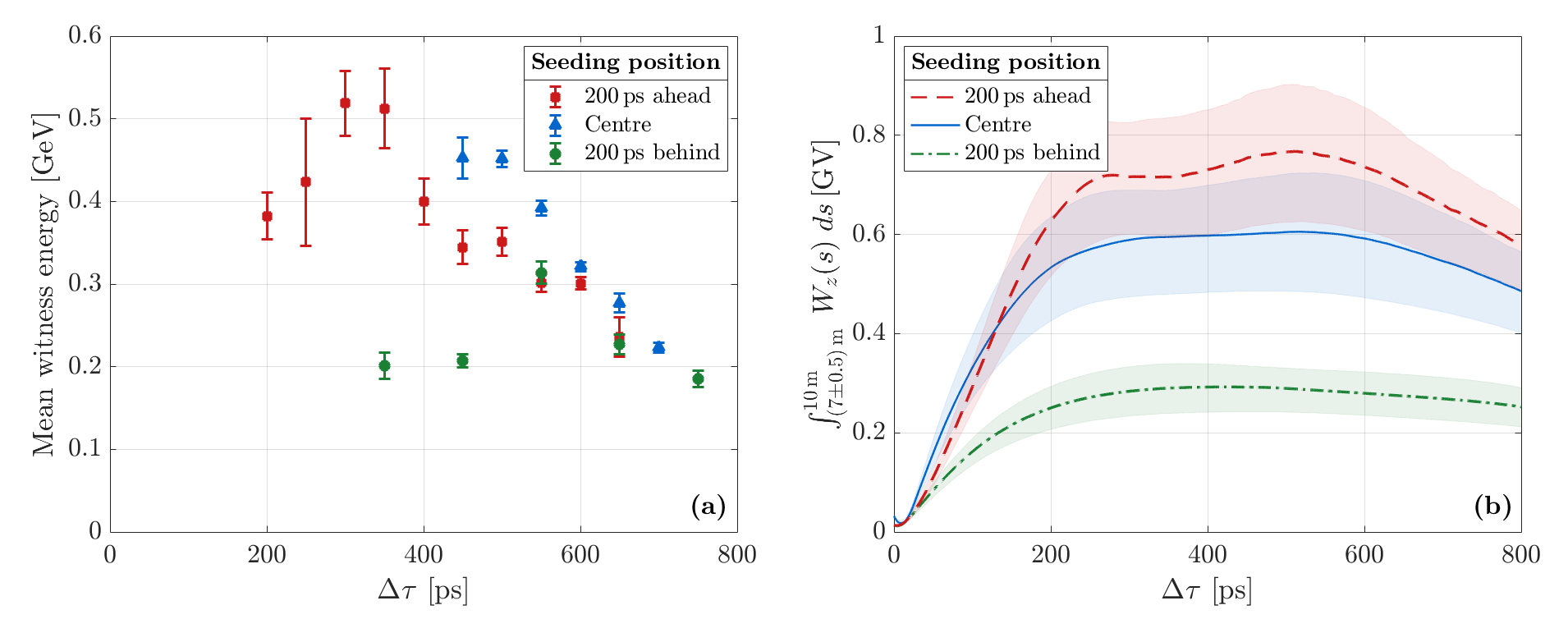}
    \caption{(a) Experimental results showing the mean witness energy with increasing laser--electron delays for the three measured seeding positions. Error bars represent the standard error on the mean. (b) Corresponding simulation results showing the envelope of the integrated longitudinal wakefield following self-modulation saturation. Error bands represent the integrated longitudinal wakefield calculated for upper and lower bounds of the estimated saturation length ${z_s = 7.0 \pm 0.5}$\,m.}
    \label{fig:longdelays}
\end{figure*}

\section{Resonant wakefield excitation}
\label{sec:resonantexcitation}

It is expected that as the laser--electron delay is increased, the energy of captured electrons increases as an increasing number of proton micro-bunches resonantly excite the wakefield that the captured witness electrons experience. Experimentally, this was found to be the case up until the laser--electron delay exceeded approximately the RMS length of the proton bunch, $\sigma_z = 257 \pm 3 \,$ps. This is demonstrated in Fig.~\ref{fig:shortdelays} for measurements when seeding 100\,ps ahead of the centre of a proton bunch of population ${\left( 3.13 \pm 0.16 \right) \times 10^{11}}$ and RMS length ${\left( 7.70 \pm 0.08\right)}$\,cm. For small delays, $\Delta \tau < 150$\,ps, achieving consistent witness capture was challenging, resulting in low detected charge with larger energy variability. In this case, approximately 30\,\% of injection attempts achieved sufficient charge capture to contribute to the measurements for each value of the laser--electron delay. However, for $\Delta \tau \geq 150$\,ps, successful witness capture was observed in more than 75\,\% of events per delay. Nevertheless, the trend for increasing witness energy with increasing laser--electron delay was observed on short timescales ($\Delta \tau < 250$\,ps) and demonstrate resonant excitation of the wakefield. At around $\Delta \tau = 250$\,ps $\approx \sigma_z$, the witness energy was observed to saturate as predicted by previous simulations~\cite{sigzmaxfield}.

\section{Extended timescale wakefield evolution}
\label{sec:longtimescale}

The data corresponding to studies of larger laser--electron delays were taken under different experimental conditions to those discussed in Sec.~\ref{sec:resonantexcitation}. The measured proton bunch parameters differed between the two measurements. The proton bunch population was measured to be ${\left( 2.83 \pm 0.14 \right) \times 10^{11}}$ while its RMS bunch length was ${\left( 7.65 \pm 0.08 \right)}$\,cm. In addition to the change in seeding position between the measurements, the reduction in bunch population while maintaining approximately the same bunch length alters the peak current of the beam. This results in a reduction in the wakefield amplitude over the plasma length and as a result, a difference in the self-modulation saturation length. Therefore, witness electrons captured in the experimental conditions discussed in Sec.~\ref{sec:resonantexcitation} will experience a larger amplitude wakefield and undergo consistent acceleration over a longer distance than those for subsequent measurements, resulting in larger witness energies. For this reason, direct comparison between the mean captured witness energy of the two measurements cannot be made. However, for the measurements studying the effect of various seeding positions on extended timescales (up to $800$\,ps) detailed in this section, the experimental parameters were kept as consistent as possible to allow direct comparison. In this case, for the smallest delays ($\Delta\tau < 200\,$ps) capture of electrons was not realised while fewer than 30\,\% of attempted events per delay achieved sufficient witness capture to contribute to the measurements of laser--electron delays between ${200-400}$\,ps. For larger delays, witness electron bunches of sufficient charge were measured in approximately 70\,\% of events per laser--electron delay. 

\subsection*{Longitudinal wakefield amplitude}

For laser--electron delays larger than 500\,ps, the effects of further plasma wakefield evolution were observed and are demonstrated in Fig.~\ref{fig:longdelays}. Experimentally, on these extended timescales a decreasing witness energy was measured with increasing laser--electron delay, indicative of a decreasing longitudinal wakefield amplitude along the self-modulated proton bunch. 

\begin{figure}[t]
	\centering
	\includegraphics[width=0.49\textwidth]{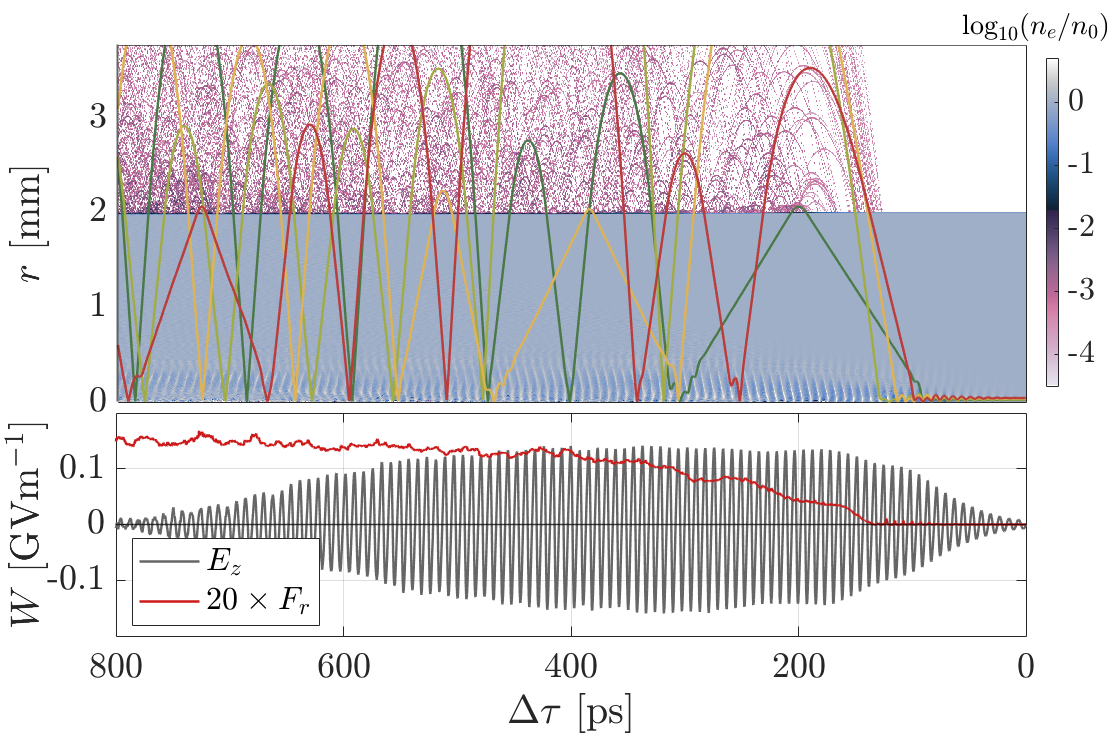}
    \caption{Upper: Simulated plasma electron density map at $z = 2.9\,$m. Electrons are ejected from the plasma as the wakefield amplitude is resonantly excited and trajectory crossing occurs near the axis. Example plasma electron trajectories are plotted for reference (solid lines). The electron density perturbation driven by the self-modulated proton bunch can be seen near the axis. Lower: Corresponding on-axis longitudinal field (grey) shows decay for large $\Delta \tau$ alongside the development of a radial field outside the plasma boundary (red), measured at $r = 7.5\,k_p^{-1}$.}
    \label{fig:etrajwavebreaking}
\end{figure}

Simulations indicate this is due to the onset of plasma electron trajectory crossing within the wakefield. The cumulative wakefield excitation drives an increasing radial electron density gradient along the proton bunch as the wakefield increases in amplitude and the plasma electrons near the axis of propagation are fully expelled. Due to the presence of the radial density gradient created by the wakefield, plasma electrons experience a spatially-dependent radial force and their initially coherent oscillations begin to mix~\cite{dawsonmixing}. This modifies the resonant frequency of the radial force while the longitudinal field continues to be excited at the initial plasma frequency~\cite{finiteplasma}. The wavefront of the density perturbation hence becomes deformed and electron trajectories start to cross. For our simulated plasma channel parameters approximately 5\,\% of plasma electrons are radially expelled from the plasma as inner electrons cross trajectories of outer electrons and experience an increased negative charge density. This cumulative effect begins at the rear of the proton bunch where the wakefield amplitude is large enough and propagates forward as the wakefield potential, and hence density perturbation, associated with each micro-bunch grows as self-modulation develops along the length of the plasma~\cite{ssm2}. After these inner electrons are expelled as demonstrated in Fig.~\ref{fig:etrajwavebreaking}, outer plasma electrons move inwards to replace them and the plasma becomes charged at its boundary~\cite{plasmacharging}. This, in combination with the presence of electrons outside the boundary of the plasma, induces a positive potential in the surrounding volume that is attracting for electrons and acts to accelerate the previously ejected electrons back towards the plasma.

As these electrons re-enter the plasma, they return back on-axis and interfere with the resonantly driven wakefield. The combination of this interference effect and the phase mixing induced by the radial density gradient damps the wakefield amplitude, causing the decay of the longitudinal wakefield observed in the lower panel of Fig.~\ref{fig:etrajwavebreaking} and the experimentally measured decrease in witness energy observed for large laser--electron delays shown in Fig.~\ref{fig:longdelays}(a). As the attracting force is low in amplitude when compared to the radial wakefield near the axis, expelled electrons re-enter the plasma far behind the position at which they are ejected from the plasma in the co-moving frame. For this reason, while trajectory crossing is observed to occur earlier than $\sigma_z$ behind the seeding position in this simulation, significant effects from the return of plasma electrons on the wakefield amplitude do not become apparent until larger laser--electron delays, $\Delta \tau \geq 500$\,ps. 

This mechanism therefore splits the evolution of the wakefield amplitude into three distinct regions along the proton bunch: (i) initially for small laser--electron delays, $\Delta \tau \leq 250\,$ps, the wakefield amplitude grows as an increasing number of proton micro-bunches contribute to driving the wakefield and plasma electron oscillations remain coherent. This is the region that is typically studied experimentally and theoretically as it offers the largest stable wakefield amplitudes (e.g.~\cite{nature, sigzmaxfield, saturationinjection}). As the proton beam undergoes self-modulation and the wakefield amplitude and density perturbation grow, trajectory crossing is observed in simulation towards the rear of this region and electrons are ejected from the plasma. (ii) At around 250\,ps behind the seeding position, the wakefield amplitude saturates. The wakefield amplitude is maintained in this region (250 $< \Delta \tau < 500$\,ps) as outer plasma electrons replace those ejected via trajectory crossing. Phase mixing effects induced by the radial density gradient cause electron oscillations to decohere and limit further growth of the wakefield amplitude despite an increasing number of micro-bunches driving the wakefield. (iii) Finally, for $\Delta \tau \geq 500$\,ps, a significant number of ejected electrons return to the plasma and their interference with the resonantly driven wakefield on-axis, in combination with accumulated phase mixing effects, cause further decay in the wakefield amplitude along the bunch. 

The appearance of plasma electron trajectory crossing so early within the resonantly excited wakefield is a consequence of the low plasma density used in these studies. At a density of $2\times10^{14}$\,cm$^{-3}$ the proton beam is narrow relative to the skin depth of the plasma, $\sigma_r \approx 200\,\mu$m $= 0.53 \,k_p^{-1}$, quickly exciting the large transverse gradients necessary to induce plasma electron trajectory crossing. At the AWAKE nominal density of $7\times10^{14}\,$cm$^{-3}$ the relative bunch density is reduced, the bunch size is optimised for the plasma density ($\sigma_r = k_p^{-1}$), and trajectory crossing does not occur. At the nominal density, it is rather expected that ion motion causes decay of the wakefield on extended timescales~\cite{ionmotion1,ionmotion2,ionmotionminakov}, an effect not observed in the simulations performed for this study.

\begin{figure*}[t]
	\centering
	\includegraphics[width=\textwidth]{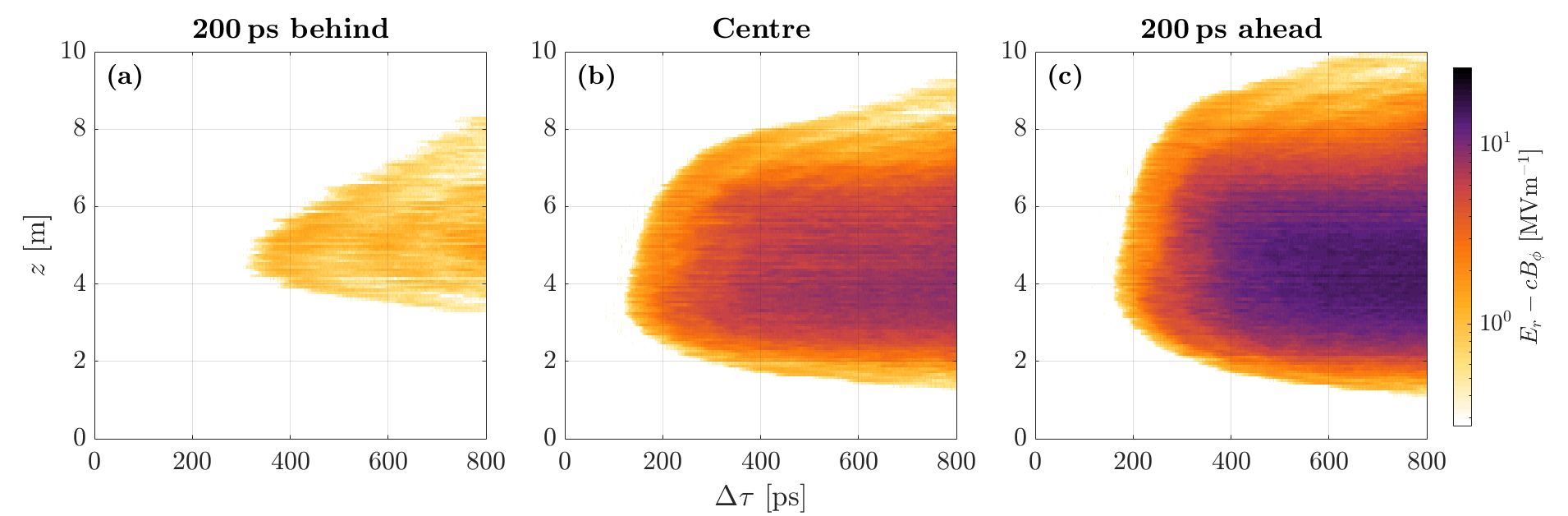}
    \caption{Simulated development of the transverse field, $E_r - c B_{\phi}$, outside the boundary of the plasma measured at $r = 7.5\,k_p^{-1}$ for the three seeding positions measured experimentally.}
    \label{fig:offaxisfield}
\end{figure*}

Figure \ref{fig:longdelays}(b) shows the envelope of the integrated longitudinal wakefield for different laser--electron delays predicted by LCODE simulations using the proton bunch parameters of the experiment. The wakefield amplitude is integrated over the plasma length following self-modulation saturation, $z_s = 7.0\,\pm\,0.5$\,m, identified in simulation by the position after which the wakefield phase remains approximately constant with respect to the proton micro-bunches. This corresponds to the position in the plasma at which consistent acceleration of witness electrons is possible. Before this point, the phase of the wakefield continuously evolves as the proton bunch self-modulates, potentially causing captured electrons to dephase and move into decelerating, defocusing regions of the wakefield~\cite{selfmodulation,saturationinjection}. The integrated wakefield amplitude shown in Fig.~\ref{fig:longdelays}(b) represents an upper bound on the expected witness energy as it assumes witness electrons remain on-axis in the position of maximal electric field amplitude over the relevant integrated plasma length. It is rather expected that witness electrons oscillate within the accelerating region of the wakefield and therefore have lower energy at the exit of the plasma than predicted by simulation, consistent with the data presented in Fig.~\ref{fig:longdelays}(a). The trends observed experimentally (Fig.~\ref{fig:longdelays}(a)) are reproduced in simulation (Fig.~\ref{fig:longdelays}(b)) with three distinct regions: initial growth of the wakefield as it is resonantly excited ($\Delta \tau < 250\,$ps), a region of field saturation ($250 \leq \Delta \tau < 550\,$ps), and decay on longer timescales. The experimental results presented in Fig.~\ref{fig:shortdelays} for a seeding position 100\,ps ahead of the centre of the proton bunch are also consistent with these findings, but show only the initial growth and saturation regions of the integrated wakefield amplitude due to the limited range of laser--electron delays measured ($\Delta \tau \leq 365$\,ps).

The magnitude of the decay of the integrated wakefield amplitude on long timescales ($\Delta \tau > 500\,$ps) shown in Fig.~\ref{fig:longdelays}(b) is dependent on the seeding position. This is expected as trajectory crossing and the subsequent ejection of plasma electrons is a direct result of the phase mixing induced by the increasing radial density gradient from the cumulative excitation of the wakefield. When seeding 200\,ps behind the centre of the proton bunch, the initial proton bunch density is lower and decreasing along the bunch when compared to the other two seeding positions. Consequently the wakefield amplitude driven by the modulated proton bunch is lower as demonstrated in Fig.~\ref{fig:beamdenscomp} and simulations indicate that trajectory crossing does not occur until later in the plasma ($z > 3$\,m) and further behind the seed position ($\Delta \tau > 300\,$ps). This is exemplified in Fig.~\ref{fig:offaxisfield} which shows the simulated development of the radial field outside the plasma boundary ($r = 7.5\,k_p^{-1}$) over the entire plasma length for the three different seeding positions tested experimentally. The peak amplitude of the potential induced outside the boundary of the plasma when seeding 200\,ps behind the centre of the proton bunch (Fig.~\ref{fig:offaxisfield}(a)) is an order of magnitude lower than for the other seeding positions, indicating far fewer electrons being ejected from the plasma. Therefore, the long timescale wakefield decay induced by the return of ejected plasma electrons is reduced for this seeding position as shown in Fig.~\ref{fig:longdelays}(b). This was similarly observed in the experimental measurements shown in Fig.~\ref{fig:longdelays}(a) where the mean witness energy is approximately constant with increasing delay when seeding 200\,ps behind the centre. 

\begin{figure*}[t]
	\centering
	\includegraphics[width=\textwidth]{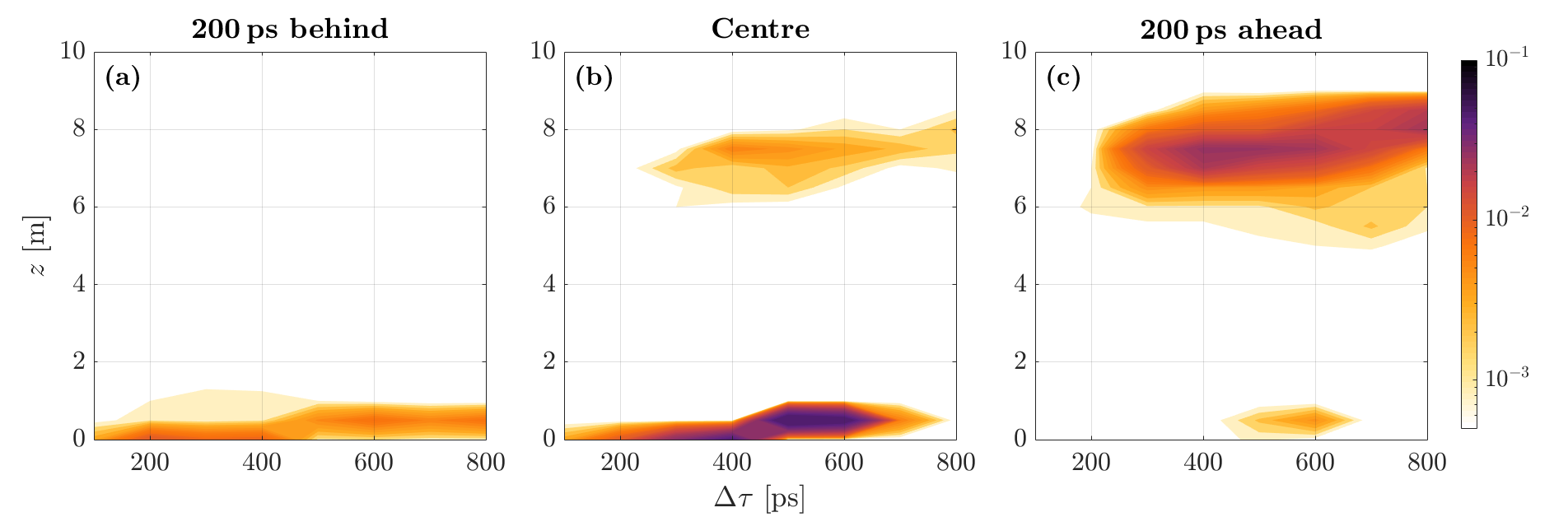}
    \caption{Maps of final injection position for witness electrons from simulations of electron injection. A separate simulation is performed for each laser--electron delay in 100\,ps steps. The colourscale represents the fraction of electrons that are captured in the wakefield at a particular $z$-position, normalised to the total number of witness electrons that are injected at the start of the simulation. A cut on the witness electron energy of $E \geq 84$\,MeV is used to mirror the experimental conditions for the spectrometer dipole.}
    \label{fig:injectionmaps}
\end{figure*}

In comparison, when seeding 200\,ps ahead of the centre of the bunch the number of protons contributing to driving the wakefield is far higher. Larger amplitude wakefields are driven earlier in the plasma (Fig.~\ref{fig:beamdenscomp}) and the transverse potential outside the plasma boundary begins to develop within $2\,$m (Fig.~\ref{fig:offaxisfield}(c)). The amplitude of the potential also exceeds that of the central seeding position shown in Fig.~\ref{fig:offaxisfield}(b). Larger on-axis wakefield amplitudes are driven when seeding 200\,ps ahead of the centre of the bunch and correspondingly more electrons undergo trajectory crossing and can later return to the plasma and damp the wakefield on-axis. This results in faster decay of the integrated wakefield amplitude for large laser--electron delays than for the central seeding position as demonstrated in Fig.~\ref{fig:longdelays}(b). 

The position of the plasma boundary, assumed to be $r_p = 2\,$mm in this study, is not measured experimentally and hence has a large uncertainty. Simulations varying the position of the boundary between reasonable experimental limits, ${r_p = 1.5 \rightarrow 2.5}$\,mm, were performed to investigate the effect of this. An $\mathcal{O}(10\%)$ change in the average amplitude of the field generated outside the plasma boundary was observed as an increased number of plasma electrons gained sufficient transverse momentum following trajectory crossing to be ejected from the plasma for a reduced plasma boundary and vice versa. However, a correspondingly smaller change in the integrated longitudinal wakefield amplitude was observed (\%-level). Therefore, over the range of positions tested in this study, the simulation results were relatively insensitive to the exact position of the plasma boundary.  

\subsection*{Witness capture}

The development of the potential outside the plasma boundary also affects the witness charge captured within the wakefield. Previous simulations of witness injection have indicated that some witness electrons injected at the entrance of the plasma can be defocused near the plasma boundary~\cite{3dpicinjection} or by the seed wakefields driven by the unmodulated proton bunch~\cite{witnessdefocussingseed}. These electrons are then observed to oscillate around or near the boundary of the plasma and continue to propagate along its length. In addition to this, while the proton bunch undergoes self-modulation the phase of the wakefield is continuously evolving with respect to both the proton micro-bunches and any witness electrons captured within the wakefield. This can cause witness electrons that have been captured at the initial injection position at the start of the plasma ($z < 1\,$m) to become dephased and move into decelerating, defocusing regions of the wakefield where they can be lost from the wakefield structure. In simulation these electrons leave the plasma quickly but are then also observed to continue to propagate along the plasma length near its boundary. In this study, simulations indicate that following the development of the potential induced outside the boundary of the plasma, witness electrons that have travelled along its boundary are re-injected into the plasma and can be captured within the wakefield. 

\begin{figure}[b]
	\centering
	\includegraphics[width=0.49\textwidth]{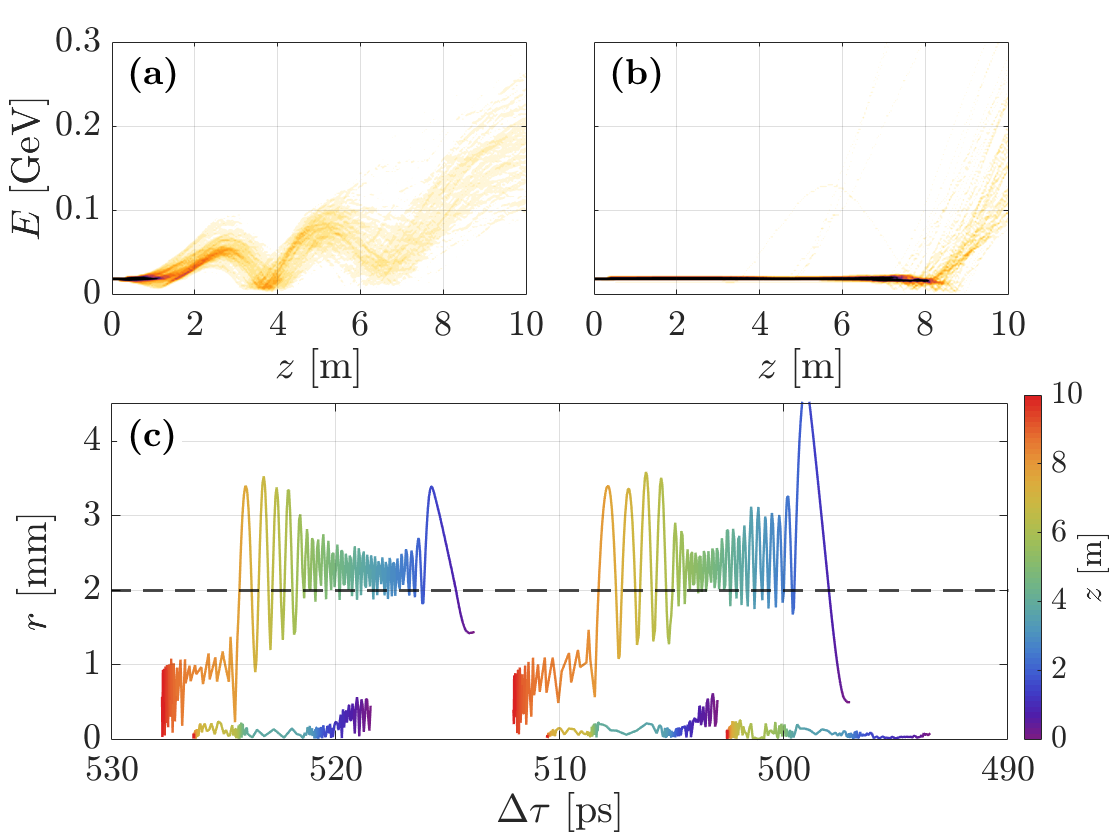}
    \caption{Simulated evolution of the witness energy for (a) electrons that remain trapped in the wakefield throughout self-modulation, and (b) electrons that are injected following development of the electron-focusing potential outside the plasma boundary. (c) Example witness electron trajectories in the co-moving frame; the colour scale represents the $z$-position of an electron. The black horizontal dashed line represents the position of the plasma boundary. This simulation corresponds to a central seeding position with $\Delta \tau = 500$\,ps.}
    \label{fig:wittraj}
\end{figure}

\begin{figure*}[t]
	\centering
	\includegraphics[width=\textwidth]{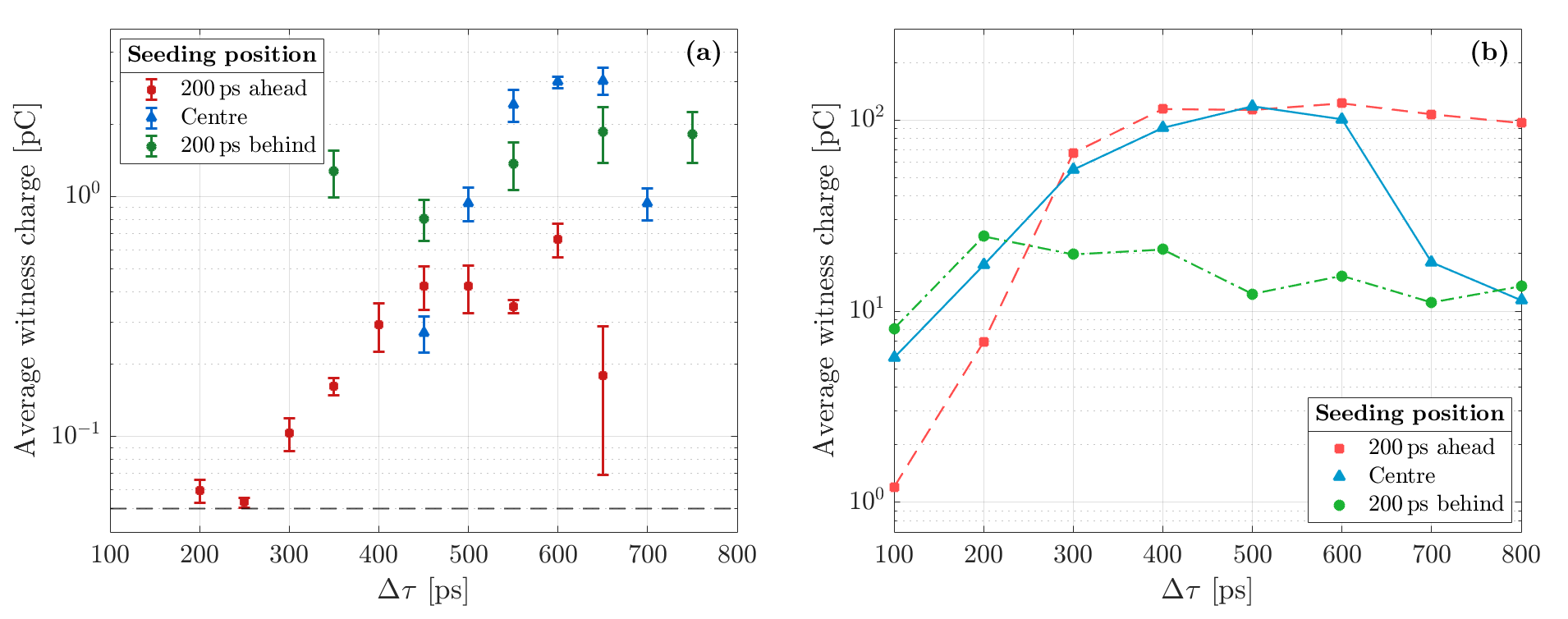}
    \caption{(a) Experimental and (b) simulation results demonstrating the mean witness charge as a function of the laser--electron delay. Trends observed experimentally are reproduced in simulation. Experimental error bars represent the standard error on the mean. The horizontal dashed line at 50\,fC represents the minimum required charge measured on the spectrometer screen for an event to be included in the analysis.}
    \label{fig:chargecapture}
\end{figure*}

The effect of this is demonstrated in Fig.~\ref{fig:injectionmaps} which shows the simulated maps of witness capture positions along the plasma for the three different seeding positions measured experimentally for variable laser--electron delays. Example witness electron trajectories are additionally shown in Fig.~\ref{fig:wittraj}(c); some electrons remain close to the axis and trapped within the wakefield over the entire plasma length, while others are observed to propagate near the plasma boundary and are injected following the development of the potential outside the plasma boundary at $z > 5\,$m.

When seeding in the centre of the proton bunch for small laser--electron delays (Fig.~\ref{fig:injectionmaps}\,(b), $\Delta \tau \leq 200\,$ps), capture only occurs at the position of the initial injection of the witness bunch at the start of the plasma ($z < 1\,$m). This is the region of the proton bunch ahead of the development of the potential outside the plasma boundary demonstrated in Fig.~\ref{fig:offaxisfield}\,(b). However, for $\Delta \tau \geq 300$\,ps a growing amount of charge is re-injected and captured in the wakefield later in the plasma ($6 \leq z \leq 9$\,m) with increasing laser--electron delay. This directly corresponds to the evolution of the electron-attracting potential outside the boundary of the plasma illustrated in Fig.~\ref{fig:offaxisfield}. Similarly, when seeding 200\,ps ahead of the centre of the proton bunch (Fig.~\ref{fig:injectionmaps}\,(c)), significant re-injection and capture is observed following the development of the potential outside the plasma boundary ($z > 5$\,m). In this case, an increased fraction is re-injected and captured when compared with seeding in the centre of the proton bunch, consistent with the larger amplitude potential induced outside the plasma boundary demonstrated in Figs.~\ref{fig:offaxisfield}\,(b) and (c). In simulation, for both seeding positions, there is not a significant difference in the final energy of witness electrons that are captured either at the start of the plasma and remain within the wakefield throughout the entire development of the self-modulation ($z < 1$\,m), or those that are re-injected and captured in the wakefield following the development of the potential outside the plasma boundary ($z > 5$\,m) as shown in Fig.~\ref{fig:wittraj}. Witness electrons that are captured in the wakefield at the start of the plasma undergo dephasing as the wakefield phase continuously evolves and are decelerated multiple times before self-modulation saturates at $z \approx 7\,$m, at which point consistent acceleration is observed. Therefore, the effective acceleration length for witness electrons captured either at the start of the plasma (Fig.~\ref{fig:wittraj}(a)) or following re-injection by the potential outside the plasma boundary (Fig.~\ref{fig:wittraj}(b)) is similar. This further motivates using the saturation position ($z_s = 7.0 \pm 0.5$\,m) as the lower limit within the calculation of the integrated wakefield in Fig.~\ref{fig:longdelays}(b) when comparing to experimental results. When seeding 200\,ps behind the centre of the proton bunch, successful capture of re-injected witness electrons late in the plasma is not observed in simulation (Fig.~\ref{fig:injectionmaps}\,(a), $z > 1\,$m). This is due to the reduced amplitude of the focusing potential outside the boundary of the plasma in this case, as demonstrated in Fig.~\ref{fig:offaxisfield}\,(a).

Experimentally, corresponding signatures consistent with the seed-dependent effect of witness re-injection following the development of a radial potential outside the plasma boundary were observed and are demonstrated in Fig.~\ref{fig:chargecapture}. This figure shows (a) the average witness charge measured experimentally and (b) observed in simulation when varying the laser--electron delay. A minimum witness energy of 84\,MeV at the exit of the plasma was applied in simulation analysis in order to mimic the spectrometer dipole settings used in the experiment. When seeding ahead of, or in, the centre of the proton bunch, an increasing witness charge was measured with increasing laser--electron delay. This is consistent with the evolution of the electron focusing field outside the boundary of the plasma created following plasma electron ejection (Fig.~\ref{fig:offaxisfield}) acting to re-inject witness electrons where they can be captured within the wakefield. However, when seeding behind the centre of the proton bunch approximately constant witness capture with increasing delay was observed both experimentally and in simulations. These experimental observations directly correspond to the trends predicted by the witness injection simulations for the three seeding positions demonstrated in Fig.~\ref{fig:chargecapture}(b). For $\Delta \tau = 700\,$ps with a central seeding position, a drop in the average witness charge is observed both experimentally and in simulation. This corresponds to a large reduction in the amount of charge captured close to the entrance of the plasma ($z < 1\,$m), as shown in Fig.~\ref{fig:injectionmaps}(b). In contrast, when seeding 200\,ps ahead of the centre of the proton bunch a smaller fraction of witness electrons are captured for $z < 1\,$m at all $\Delta \tau$, resulting in a more consistent average witness charge at large $\Delta \tau$ both experimentally and in simulation as the re-injection mechanism dominates. 

As expected, the simulations predict average captured witness charges that are more than an order of magnitude larger than those measured experimentally for geometric reasons discussed previously. While the development of this re-injection mechanism appears to offer a method for increasing charge capture experimentally, it is important to note that this secondary injection is not controlled. It will not result in high quality witness bunches and as such is not desirable. 

The injection scheme for AWAKE Run 2 differs from that used in Run 1 in order to improve capture, preserve emittance and minimise the energy spread of the injected witness bunch by injecting the bunch following self-modulation saturation~\cite{run2patric}.  This will minimise sources of the secondary injection mechanism identified in this study that can degrade the witness bunch quality. In addition to this, the largest acceleration gradients were observed ahead of the onset of effects created by trajectory crossing and as such the long timescale wakefield effects explored here are not expected to limit the accelerating gradient, energy gain or charge capture for AWAKE Run 2. 

\section{Conclusion}

The evolution of the amplitude of the wakefield driven along a self-modulated proton bunch in plasma is measured experimentally by varying the relative timing of the wakefield seeding position and the injection position of witness electrons. The use of a low plasma density ($n_e = 2\times10^{14}$\,cm$^{-3}$) and a correspondingly narrow drive bunch ($\sigma_r = 0.53\,k_p^{-1}$) results in the fast development of strong radial transverse plasma density gradients. The presence of radial gradients causes phase mixing and plasma electron trajectory crossing within the wakefield, leading to the expulsion of plasma electrons. An electron-focusing potential is induced in the volume outside the plasma boundary and acts to accelerate electrons back into the plasma. The return of electrons damps the wakefield amplitude while increasing witness capture. The seeding position is changed to investigate the effect of the longitudinal wakefield amplitude on the development of the electron-focusing potential outside the plasma boundary showing amplified effects for higher amplitude wakefields and vice versa. Agreement between the witness energy and charge capture trends observed both experimentally and in simulation provides evidence that the mechanisms identified in this study are the cause of the decrease in energy gain and increase in charge capture measured for large laser--electron delays. This study therefore contributes useful information for optimisation of the acceleration process for AWAKE Run 2 and beyond.

It is expected that the occurrence of trajectory crossing within the resonantly excited wakefield is a result of using a narrow drive beam and hence should be suppressed at higher plasma densities where the relative transverse beam size is larger. Therefore, to further study the development of this mechanism, similar measurements could be repeated at a range of plasma densities. This would allow determination of the relative transverse beam size at which the experimental signatures associated with trajectory crossing can be observed, thus providing a useful cross-check with simulation predictions and helping to optimise beam and plasma parameters for future applications. 

\section*{Acknowledgements}
This work was supported in parts by a Leverhulme Trust Research Project Grant RPG-2017-143 and by STFC (AWAKE-UK, Cockcroft Institute core, John Adams Institute core, and UCL consolidated grants), United Kingdom; a Deutsche Forschungsgemeinschaft project grant PU 213-6/1 ``Three-dimensional quasi-static simulations of beam self-modulation for plasma wakefield acceleration''; the National Research Foundation of Korea (Nos. NRF-2016R1A5A1013277 and NRF-2019R1F1A1062377); the Portuguese FCT---Foundation for Science and Technology, through grants CERN/FIS-TEC/0032/2017, PTDC-FIS-PLA-2940-2014, UID/FIS/50010/2013 and SFRH/IF/01635/2015; NSERC and CNRC for TRIUMF's contribution; the Research Council of Norway; the Wolfgang Gentner Programme of the German Federal Ministry of Education and Research (grant no. 05E15CHA); and the U.S. National Science Foundation under grant PHY-1903316. M. Wing acknowledges the support of the Alexander von Humboldt Stiftung and DESY, Hamburg. Contribution of the Novosibirsk team was supported by the Russian Science Foundation, project 20-12-00062. Support of the Wigner Datacenter Cloud facility through the "Awakelaser" project is acknowledged. The work of V. Hafych has been supported by the European Union's Framework Programme for Research and Innovation Horizon 2020 (2014--2020) under the Marie Sklodowska-Curie Grant Agreement No.765710. The authors acknowledge the use of the UCL Myriad and Grace High Performance Computing Facilities (Myriad@UCL, Grace@UCL), and associated support services, in the completion of this work. The AWAKE collaboration acknowledge the SPS team for their excellent proton delivery.


\end{document}